\documentstyle[preprint,eqsecnum,aps]{revtex}
\begin{document}
\draft
\title{Existence of the Wigner function with correct marginal
distributions along tilted lines on a lattice}
\author{Minoru HORIBE, Akiyoshi TAKAMI, Takaaki HASHIMOTO and Akihisa HAYASHI}
\address{
Department of Applied Physics, Fukui University, Fukui 910, Japan
}
\date{\today}
\maketitle
\begin{abstract}
In order to determine the Wigner function uniquely, we
introduce a new condition which ensures that the Wigner function
has correct marginal distributions along tilted lines.
For a system in $N$ dimensional Hilbert space,
whose ``phase space" is a lattice with $N^2$ sites,
we get different results depending on whether $N$ is odd or even.
Under the new condition, the Wigner function is determined if $N$ is an
odd number, but
it does not exist if $N$ is even.
\end{abstract}
\section{Introduction}
In the quantum theory, we can define the ``weight", although this
weight may be negative, and the expectation
value of an observable is to be expressed as the average
of a real function on phase space corresponding to 
the observable. This weight is well-known as the Wigner
function\cite{wigner} and usually assumed to satisfy
the conditions:

\vspace{3mm}

\noindent
(a) We can obtain the marginal distributions along the coordinate
and momentum axes by integration.\\
(b) The Wigner function is real valued. \\
(c) The Wigner function includes the same information as the density
matrix.\\

\vspace{3mm}

\noindent For a continuous quantum system
with the coordinate $\hat{Q}$ and the momentum $\hat{P}$,
these conditions are described in the following forms:

\vspace{3mm}

\noindent (a)The integration of the Wigner function $W(q,p)$ over the
variable $p$ gives the expectation value of the density
matrix $\hat{\rho}$ for the eigenvector $|q \rangle$ of $\hat{Q}$
\[
\int_{-\infty}^{\infty}W(q,p)dp = \langle q |\hat{\rho} | q \rangle.
\]
Similarly, the expectation value of the density matrix $\hat{\rho}$
for the eigenvector $|p \rangle$ of $\hat{P}$ is obtained
by integration of the Wigner function over the variable $q$
\[
\int_{-\infty}^{\infty}W(q,p)dq = \langle p |\hat{\rho} | p \rangle.
\] 
(b)The Wigner function is real valued, 
\[
W^{\ast}(q,p)=W(q,p).
\]
\noindent (c) Usually this condition is written in terms of the Fano operator\cite{fano} which
is defined by
\begin{equation}
W(q,p)={\rm Tr}[\hat{\Delta}(q,p)\hat{\rho}].
\label{den2wig}
\end{equation}
Moreover it is imposed that the Hermite conjugate operator $\hat{\Delta}^{\dagger}(q,p)$ 
is employed for the inverse relation
\begin{equation}
\hat{\rho}=2\pi \hbar\int_{-\infty}^{\infty}dq dp
\hat{\Delta}^{\dagger}(q,p)W(q,p),
\label{wig2den}
\end{equation}
although these equations are not equivalent to condition (c)
but sufficient conditions for it.
It is not difficult to show the orthogonality and completeness
of the Fano operator.

We can check that these conditions are satisfied by the
function $W(q,p)$
\begin{equation}
W(q,p)=\frac{1}{2 \pi \hbar}\int_{-\infty}^{\infty}dr
  \left[
      e^{-ipr/\hbar} 
      \left\langle q+\frac{r}{2} \right | \hat{\rho} 
      \left|       q-\frac{r}{2} \right\rangle
   \right].
\label{orgwigc}
\end{equation}
For this Wigner function the Fano operator becomes
\begin{equation}
\hat{\Delta}(q,p)=\frac{1}{2\pi \hbar}\int_{-\infty}^{\infty} dr
\left|q-\frac{r}{2} \right\rangle
e^{-ipr/\hbar} 
\left\langle q+\frac{r}{2} \right|.
\label{orgfanoc}
\end{equation}

Inversely, when we attempt to determine the Wigner function
under these conditions we find ambiguity
as is pointed out by many physicists 
\cite{wigner2}-\cite{tatarskii} and
in a previous paper\cite{hashimoto}. A new condition is needed for the
determination of the Wigner function. Bertrand and Bertrand
\cite{BertTomo} essentially required correct marginals along
tilted lines in addition to basic
marginals ({\em i.e.} condition (a))
and showed that there exists only one solution Eq.(\ref{orgwigc}).

This condition is important not only for the determination of
the Wigner function but also for quantum tomography.
When we can get the marginal distributions on all tilted
lines from observations, using
the inverse Radon transformation we find out the Wigner function which has the same information
as the density matrix. Specially, if we consider a pure state, we can obtain the wave function
up to a phase factor from the observations.

Now we consider a quantum system in a vector space spanned by $N$ eigenvectors 
$\mid q \rangle \;\;(q=0,1 \cdots ,N-1)$ for the ``coordinate" operator. 
It is assumed that these eigenvectors satisfy periodical boundary conditions
\begin{equation}
\mid q+N \rangle = \mid q \rangle .
\label{pbc}
\end{equation}
The eigenstates of the ``momentum" operator are defined by
\begin{equation}
\mid p \rangle = \frac{1}{\sqrt{N}}
\sum_{q \in Z_N}\omega^{-qp} \mid q \rangle ,
\label{defmeive}
\end{equation}
and the phase space of this system is a lattice with $N^2$ sites.
Here, $\omega$ is a primitive $N$th root of unity
\[
\omega=e^{i\frac{2\pi}{N}}.
\]
When the conditions corresponding to conditions (a) $\sim$ (c)
are imposed, Cohendet {\em et al.}\cite{cohendet}
gave the Wigner function for the case where $N$ is odd.
For the case with $N$=even number, we presented
the Fano operator in the previous paper\cite{hashimoto}.
 These solutions are not unique and we showed that the general solution
under the three conditions is obtained by multiplying
an $(N-1)^2 \times (N-1)^2$ orthogonal matrix
to a special solution\cite{hashimoto}.
Leonhardt permitted arguments of the Fano operator
to take values of half-integer besides integer values and found
the Fano operator for $N$=even number which
satisfies the three conditions corresponding to conditions (a) $\sim$ (c)
\cite{leonhardt}.
However, we assume the arguments of the Fano operator to take only integer, namely,
eigenvalues of coordinate and momentum operators in this note.
Wootters discussed the Wigner function for the case
where $N$ is prime and found 
the function which satisfies the property similar to
the marginalization on tilted lines
for continuous systems\cite{wootters}.

In this note, first, for a continuous quantum system we propose a
condition determining the Wigner function uniquely which is
essentially equivalent to that of
the Bertrand and Bertrand\cite{BertTomo} but which is in a simpler
form than they imposed. We easily extend this condition to the system
with a lattice phase space.  
In both cases,  we can see that the Wigner function under
the new condition has correct
marginal distributions along tilted lines in phase space.

\section{the Wigner function in a continuous quantum system}
\subsection{Behavior of the Fano operator under the rotation}
In this section we discuss the Wigner function for
a continuous quantum system
paying attention to the marginal distribution along tilted lines.
In the classical theory, we can regard any tilted line in phase
space $(q,p)$ as the axis for new
variable $q'$ defined by
\[
\left\{\begin{array}{l}
          q'=q \cos\theta-p \sin\theta \\
          p'=q \sin\theta+p \cos\theta
        \end{array} \right..
\]
Given the distribution function $\rho(q,p)$ for old variables,
the distribution $\rho'(q',p')$
for new variables is obtained from the simple transformation rule 
\begin{equation}
\rho'(q',p')=\rho(q,p)
=\rho(q'\cos\theta +p'\sin\theta,-q'\sin\theta + p'\cos\theta).
\label{clatradis}
\end{equation}
In the quantum theory, the transformation corresponding to the above
linear transformation is generated by the unitary operator $R_{\theta}$
\begin{equation}
R_{\theta}=e^{-i\frac{\theta}{2}
( \hat{Q}^2 + \hat{P}^2)/\hbar},
\label{rotunit}
\end{equation}
because we have
\begin{eqnarray}
\hat{Q}&\rightarrow&\hat{Q}' \equiv 
R_{\theta}\hat{Q}R_{\theta}^{-1} 
=\hat{Q}\cos\theta - \hat{P}\sin\theta, \\
\hat{P}&\rightarrow&\hat{P}' \equiv 
R_{\theta}\hat{P}R_{\theta}^{-1} 
=\hat{P} \cos\theta + \hat{Q} \sin\theta. 
\label{tranfqp}
\end{eqnarray}
Under this transformation, the Fano operator $\hat{\Delta}(q,p)$ is transformed to
$\hat{\Delta}'(q,p)$
following the same rule as usual:
\[
\hat{\Delta}(q,p) \rightarrow \hat{\Delta}'(q,p)=R_{\theta}\hat{\Delta}(q,p)R_{\theta}^{-1}.
\]
From Eq.(\ref{den2wig}), the Wigner function $W'(q',p')$ of the same density matrix
$\hat{\rho}$ is given by
\[
W'(q',p')={\rm Tr}[\hat{\rho} R_{\theta}\hat{\Delta}(q',p')R_{\theta}^{-1}].
\]
Since we consider the Wigner function as the distribution function in the classical theory,  
we require that the Wigner function has the same property as Eq.(\ref{clatradis}):
\[
W'(q',p')=W( q'\cos\theta+ p'\sin\theta,- q'\sin\theta+ p'\cos\theta).
\]
Thus we have the new condition
\begin{equation}
R_{\theta}\hat{\Delta}(q,p)R_{\theta}^{-1}
=\hat{\Delta}( q\cos\theta+ p\sin\theta,- q\sin\theta+ p\cos\theta).
\label{newconc}
\end{equation}

We can prove that the Fano operator (\ref{orgfanoc}) satisfies this condition
by using the Fano operator in the form
\begin{equation}
\hat{\Delta}(q,p)
=\frac{1}{(2\pi \hbar)^2}
\int_{-\infty}^{\infty}dsdt
e^{i\{t(\hat{P}-p)-s(\hat{Q}-q)\}/\hbar},
\label{fanoanth}
\end{equation}

Inversely, can we determine the Winger function namely the Fano operator from
these four conditions (a) $\sim$ (c) and Eq.(\ref{newconc}) ?
In the following subsections, we discuss this point.
\subsection{the Fano operator under the new condition}
We start with rewriting conditions (a) $\sim$ (c) in terms of the Fano operator.
Using the relations Eq.(\ref{den2wig}) and Eq.(\ref{wig2den})
between the Wigner function and the Fano operator
we get the following equations:
\begin{eqnarray}
{\rm (a)}&:&\int^{\infty}_{-\infty}\hat{\Delta}(q,p)dp=\mid q \rangle \langle q \mid, 
\label{conA1f} \\
           &&\int^{\infty}_{-\infty}\hat{\Delta}(q,p)dq =\mid p \rangle \langle p \mid, 
\label{conA2f} \\
{\rm (b)}&:& \hat{\Delta}^{\dag}(q,p)=\hat{\Delta}(q,p),
\label{conBf} \\
{\rm (c)}&:&\int^{\infty}_{-\infty}
\langle q_1  \mid \hat{\Delta}^{\dag}(q,p) \mid q_2 \rangle
\langle q_1' \mid \hat{\Delta}(q,p) \mid q_2' \rangle dqdp
=\delta(q_1-q'_2)\delta(q'_1-q_2),
\label{conC1f} \\
           &&{\rm Tr}[\hat{\Delta}(q,p) \hat{\Delta}^{\dag}(q',p')]=\delta(q-q')\delta(p-p').
\label{conC2f}
\end{eqnarray} 
Because of completeness of the operators
$e^{i{\cal Q}\hat{P}/\hbar}$ and 
$e^{-i\hat{Q}{\cal P}/\hbar}$ 
, we can expand the Fano operator with these operators,
\begin{equation}
\hat{\Delta}(q,p)=\frac{1}{2\pi \hbar}
\int_{-\infty}^{\infty}d{\cal Q}d{\cal P}
a(q,p;{\cal Q},{\cal P})
e^{i{\cal Q}\hat{P}/\hbar}e^{-i\hat{Q}{\cal P}/\hbar}.
\label{fanoexp}
\end{equation}
This expansion makes it easy to calculate the transformed
Fano operator by the unitary operator $R_{\theta}$.
In terms of the Fourier transformed coefficient $\tilde{a}(s,t;{\cal Q},{\cal P})$ which are defined by
\begin{equation}
\tilde{a}(s,t;{\cal Q},{\cal P})=\frac{1}{2\pi \hbar}
\int_{-\infty}^{\infty}\left[
e^{-iqs/\hbar}e^{ipt/\hbar}a(q,p;{\cal Q},{\cal P})
dqdp \right],
\label{afurier}
\end{equation}
three conditions (a) $\sim$ (c) are described in the simple forms
\begin{eqnarray}
{\rm (a)}:\:\:&& \tilde{a}(s,0;{\cal Q},{\cal P})=\delta({\cal Q})\delta({\cal P}-s), 
\label{cconA}\\
&& \tilde{a}(0,t;{\cal Q},{\cal P})=\delta(t-{\cal Q})\delta({\cal P}),
\label{cconB}\\
{\rm (b)}:\:\:&& \tilde{a}(s,t;{\cal Q},{\cal P})^{\ast}=e^{-i{\cal Q}{\cal P}/\hbar}
\tilde{a}(-s,-t;-{\cal Q},-{\cal P}),
\label{cconC} \\
{\rm (c)}:\:\:&& \int_{-\infty}^{\infty}ds dt 
\;[\tilde{a}(s,t;{\cal Q},{\cal P})^{\ast}\tilde{a}(s,t;{\cal Q}',{\cal P}') ]
\nonumber   \\
  && \;\;\;\;\;\;\;\;\;\;\;\;\;
=\delta({\cal Q}-{\cal Q}')\delta({\cal P}-{\cal P}'),
\nonumber \\
&& \int_{-\infty}^{\infty}d{\cal Q} d{\cal P} 
\;[\tilde{a}^{\ast}(s,t;{\cal Q},{\cal P})
   \tilde{a}(s',t';{\cal Q},{\cal P}) ]
\nonumber   \\
  && \;\;\;\;\;\;\;\;\;\;\;\;\;
=\delta(s-s')\delta(t-t').
\label{cconD}
\end{eqnarray}
Since the operators $e^{-i\hat{Q}{\cal P}/\hbar}$ and
$e^{i{\cal Q}\hat{P}/\hbar}$  are changed to
\begin{eqnarray}
R_{\theta}e^{-i\hat{Q}{\cal P}/\hbar}R^{-1}_{\theta}
&=& e^{-i\cos\theta\sin\theta{\cal P}^2/2}
e^{-i\cos\theta \hat{Q}{\cal P}/\hbar}
e^{ i\sin\theta \hat{P}{\cal P}/\hbar},
\label{TraeQ} \\
R_{\theta}e^{ i{\cal Q}\hat{P}/\hbar}R^{-1}_{\theta}
&=& e^{i\cos\theta\sin\theta{\cal Q}^2/2}
e^{ i{\cal Q}\sin\theta \hat{Q}/\hbar}
e^{ i{\cal Q}\cos\theta \hat{P}/\hbar},
\label{TraeP}
\end{eqnarray}
under the transformation by $R_{\theta}$,
we get the condition for $\tilde{a}(s,t;{\cal Q},{\cal P})$
from the new condition Eq.(\ref{newconc}), 
\begin{eqnarray}
&&\tilde{a}(s,t\,;\,{\cal Q},{\cal P})e^{-i{\cal Q}{\cal P}/2\hbar} \nonumber\\
&=&\tilde{a}( s\cos\theta+ t\sin\theta, t\cos\theta- s\sin\theta\,;\,
 {\cal Q}\cos\theta- {\cal P}\sin\theta,{\cal Q}\sin\theta + {\cal P}\cos\theta) \nonumber \\ 
&&\times e^{-i( {\cal Q}\cos\theta- {\cal P}\sin\theta)
({\cal Q}\sin\theta+{\cal P}\cos\theta)/2\hbar}.
\label{ncrelta}
\end{eqnarray} 
For each point $(s,t)$, we can choose the angle $\theta$ such that $s\cos\theta+ t\sin\theta = 0 $. 
\begin{eqnarray*}
\cos\theta &=& \frac{t}{\sqrt{s^2+t^2}}, \\
\sin\theta &=&-\frac{s}{\sqrt{s^2+t^2}}. 
\end{eqnarray*}
Then, from Eq.(\ref{ncrelta}), we obtain
\begin{eqnarray*}
\tilde{a}(s,t\,;\,{\cal Q},{\cal P})e^{-i{\cal Q}{\cal P}/2\hbar}
&=&\tilde{a}\left(
0,\sqrt{s^2+t^2}\,;\,
\frac{ t{\cal Q}+s{\cal P}}{\sqrt{s^2+t^2}},
\frac{-s{\cal Q}+t{\cal P}}{\sqrt{s^2+t^2}}
\right) \\
&&\;\;\; \times
\exp\left[-i\frac{( t{\cal Q}+s{\cal P})(-s{\cal Q}+t{\cal P})}
                {2\hbar (s^2+t^2)}
\right].
\end{eqnarray*} 
Using condition (a) and the property of the $\delta$-function,
we can rewrite the right hand side of the 
above equation in a simple form and obtain 
\[
\tilde{a}(s,t;{\cal Q},{\cal P})e^{-i{\cal Q}{\cal P}/2\hbar}
=\delta({\cal Q}-t)\delta({\cal P}-s),
\]
which is equivalent to the coefficient for the original
Fano operator given by (\ref{orgfanoc}). 

Thus, under the new condition Eq.(\ref{newconc})
we can determine the Fano operator 
uniquely. 

In this section, we have investigated the behavior of
the Fano operator under
the rotations induced by the unitary operator $R_{\theta}$
in phase space. We have to extend the rotations to more
general transformations in order to apply this method to the Wigner function
on a lattice, because it is impossible to find a rotation of general angles
which have one-to-one correspondence on lattice sites. Instead of the unitary
operator $R_{\theta}$, when we apply the unitary operator $R_{\alpha \beta}$
\[
R_{\alpha\beta}=e^{-i(\alpha \hat{Q}^2 + \beta \hat{P}^2)/\hbar},
\]
we have the transformation from the operators $\hat{Q}$ and $\hat{P}$
to the operator $\hat{Q}_G$ and $\hat{P}_G$,
\begin{eqnarray*}
\hat{Q}&\rightarrow&\hat{Q}_{G} \equiv 
R_{\alpha\beta}\hat{Q}R_{\alpha\beta}^{-1} 
=A\hat{Q}  +B \hat{P}, \\
\hat{P}&\rightarrow&\hat{P}_{G} \equiv 
R_{\alpha\beta}\hat{P}R_{\alpha\beta}^{-1} 
=D \hat{P} +C\hat{Q}, 
\end{eqnarray*}
so that we get
\begin{eqnarray}
R_{\alpha \beta}e^{-i\hat{Q}{\cal P}/\hbar}R^{-1}_{\alpha \beta}
&=& e^{iAB {\cal P}^2/2}
e^{-iA \hat{Q}{\cal P}/\hbar}e^{-iB \hat{P}{\cal P}/\hbar},
\label{gTraeQ} \\
R_{\alpha \beta}e^{ i{\cal Q}\hat{P}/\hbar}R^{-1}_{\alpha \beta}
&=& e^{iCD{\cal Q}^2/2}
e^{ i{\cal Q}C \hat{Q}/\hbar}e^{ i{\cal Q}D \hat{P}/\hbar},
\label{gTraeP}
\end{eqnarray}
where coefficients $A$,$B$,$C$ and $D$ are
\begin{eqnarray*}
A &=& \cos(2\sqrt{\alpha\beta}), \\
B &=& -\sqrt{\frac{\beta}{\alpha}}\sin(2\sqrt{\alpha\beta}), \\
C &=& \sqrt{\frac{\alpha}{\beta}}\sin(2\sqrt{\alpha\beta}), \\
D &=& \cos(2\sqrt{\alpha\beta}).
\end{eqnarray*}
After the almost same calculations as the above, we can show that
the Fano operator in Eq.(\ref{orgfanoc}) or Eq.(\ref{fanoanth}) satisfies the
similar relation
\begin{equation}
R_{\alpha\beta}\hat{\Delta}(q,p)R_{\alpha\beta}^{-1}
=\hat{\Delta}(Dq-Bp,-Cq+Dp).
\label{gnewconc}
\end{equation}
If we can find the linear transformation with one-to-one correspondence among lattice sites,
we can do the same thing on a lattice.

In the next section we try to apply this method to finding
the Fano operator on a lattice. 
\section{the Wigner function on a lattice}
For a system with a lattice phase space explained in Introduction, 
we play the same game as we did in the preceding section.

We impose the following conditions 
${\rm (a)'}$ $\sim$ ${\rm (c)'}$ on the Fano operator 
$\hat{\Delta}(q,p)\;(q,p \in Z_N)$ 
\begin{eqnarray}
{\rm (a)'}:\:\:&& \sum_{p \in Z_N} \hat{\Delta}(q,p)=
       | q   \rangle \langle q  |, \label{conda} \\
              && \sum_{q \in Z_N} \hat{\Delta}(q,p)=
       | p   \rangle \langle p  |,  \label{condb} \\
{\rm (b)'}:\:\:&& \hat{\Delta}(q,p)^{\dagger}=\Delta(q,p), \label{condc} \\
{\rm (c)'}:\:\:&&
\sum_{p,q \in Z_N}\langle q_1|\hat{\Delta}(q,p)^{\dagger}| q_2 \rangle
                  \langle q'_1|\hat{\Delta(q,p)}| q'_2 \rangle
=\frac{1}{N}\delta^{(N)}_{q_1,q'_2}\delta^{(N)}_{q_2,q'_1},
       \nonumber \\       
&&{\rm Tr}[\hat{\Delta}(q,p)\hat{\Delta}(q',p')^{\dagger}]
=\frac{1}{N}\delta^{(N)}_{q,q'}\delta^{(N)}_{p,p'},
\label{condd}
\end{eqnarray}
which correspond to conditions (a) $\sim$ (c) on the Fano operator
in a continuous quantum system.
Here, $\delta^{(N)}_{q,q'}$ is the Kronecker delta on $Z_N$,
\[
\delta^{(N)}_{q,q'}=\left\{
\begin{array}{cc}
1 & (q = q'\;\; {\rm mod}\;\; N) \\
0 & (q \neq q'\;\;{\rm mod}\;\; N) 
\end{array} \right..
\]

Owing to the expansion of the Fano operator by the operators 
$e^{i{\cal Q}\hat{P}/\hbar}$ and $e^{-i\hat{Q}{\cal P}/\hbar}$, we 
wrote some properties in simple forms in the preceding  section
for a continuous quantum system.
We introduce the matrices $P$ and $S$ corresponding to these
operators. The eigenvectors for the matrix $P$ are
$\mid q \rangle \;(q=0,1, \cdots N-1)$,
\begin{equation}
P\mid q \rangle = \omega^q \mid q \rangle,
\label{eigenp}
\end{equation}
and the eigenvectors for the matrix $S$ is
$\mid p \rangle \;(p=0,1, \cdots N-1)$ which is defined
by Eq.(\ref{defmeive})
\begin{equation}
S\mid p \rangle = \omega^{-p} \mid p \rangle.
\label{eigens}
\end{equation}
In the representation where the matrix $P$ is diagonalized,
we can write the matrices $P$ and $S$ explicitly
\begin{eqnarray}
P&=&\left(\begin{array}{ccccc}
              1   & 0      & \cdots   & \cdots &   0    \\
              0   & \ddots &          &        & \vdots \\
           \vdots &        & \omega^n &        & \vdots \\
           \vdots &        &          & \ddots &   0    \\
              0   &        &          &   0    &  \omega^{(N-1)}
        \end{array} \right),
\label{phase} \\
S&=&\left(\begin{array}{cccc}
              0   & 1      & \cdots &   0    \\
           \vdots & \ddots & \ddots & \vdots \\
           \vdots &        & \ddots &   1    \\
              1   &        &        &   0
        \end{array} \right).
\label{shift}
\end{eqnarray}
Using this representation, we can check that these matrices
satisfy the commutation property
\begin{equation}
SP=\omega PS,
\label{commSP}
\end{equation}
which is similar to the commutation property 
\[
e^{-i\hat{Q}{\cal P}/\hbar}e^{i{\cal Q}\hat{P}/\hbar}
=e^{-i{\cal Q}{\cal P}/\hbar}
e^{-i{\cal Q}\hat{P}/\hbar}e^{-i\hat{Q}{\cal P}/\hbar}.
\]
We can consider matrices $P^{\cal P}$
and $S^{\cal Q}$  on the lattice as the operators 
$e^{-i\hat{Q}{\cal P}/\hbar}$ and $e^{i{\cal Q}\hat{P}/\hbar}$,
respectively. We expect that the same arguments as in the preceding section
holds in this case.
We expand the Fano operator in power series of matrices $S$ and $P$;
\begin{equation}
\hat{\Delta}(q,p)=\sum_{n,m \in Z_N}a(q,p;n,m)S^nP^m.
\label{expdFano}
\end{equation}
Using these coefficients $a(q,p;n,m)$, conditions ${\rm (a)'}$ $\sim$
${\rm (c)'}$ become
\begin{eqnarray}
{\rm (a)'}:\;\;&&\sum_{p \in Z_N}a(q,p;n,m)
=\frac{1}{N}\omega^{-qm}\delta^{(N)}_{n,0},
\label{finF1} \\
&&\sum_{q \in Z_N}a(q,p,;n,m)=\frac{1}{N}\omega^{pn},
\label{finF2} \\
{\rm (b)'}:\;\;&& a(q,p;n,m)=\omega^{-nm}a^{\ast}(q,p;N-n,N-m),
\label{finF3} \\
{\rm (c)'}:\;\;&& \sum_{q,p \in Z_N}a^{\ast}(q,p;n,m)a(q,p;k,l)=\frac{1}{N^2}
             \delta^{(N)}_{n,k}\delta^{(N)}_{m,l},\nonumber \\
&& \sum_{k,l \in Z_N}a^{\ast}(q,p;k,l)a(q',p';k,l)=\frac{1}{N^2}
             \delta^{(N)}_{q,q'}\delta^{(N)}_{p,p'}.
\label{finF4}
\end{eqnarray}
Introducing the Fourier transformed coefficient $\tilde{a}(s,t;n,m)$
\begin{equation}
\tilde{a}(s,t;n,m)=\frac{1}{N^2}
             \sum_{q,p \in Z_N}\omega^{qs}\omega^{-pt}a(q,p;n,m),
\label{IfurierA}
\end{equation}
we can rewrite the above conditions in the simple forms:
\begin{eqnarray}
{\rm (a)'}:\;\;&&
 \tilde{a}(s,0;n,m)=\frac{1}{N^2}\delta^{(N)}_{n,0}\delta^{(N)}_{m,s}, 
\label{tfinF1}\\
&&
 \tilde{a}(0,t;n,m)=\frac{1}{N^2}\delta^{(N)}_{m,0}\delta^{(N)}_{n,t},
\label{tfinF2}\\
{\rm (b)'}:\;\;&&
 \tilde{a}(s,t;n,m)=\omega^{-nm}\tilde{a}^{\ast}(N-s,N-t;N-n,N-m),
\label{tfinF3}\\
{\rm (c)'}:\;\;&&
 \sum_{s ,t  \in Z_N}\tilde{a}^{\ast}(s,t;n,m)\tilde{a}(s,t;k,l)
                     =\frac{1}{N^4}\delta^{(N)}_{n,k}\delta^{(N)}_{m,l},\nonumber\\
&&
 \sum_{k ,l  \in Z_N}\tilde{a}^{\ast}(s,t;k,l)\tilde{a}(s',t';k,l)
                     =\frac{1}{N^4}\delta^{(N)}_{s,s'}\delta^{(N)}_{t,t'}.
\label{tfinF4}
\end{eqnarray}
In the previous paper\cite{hashimoto}, we showed that there exist 
many matrices satisfying conditions ${\rm (a)'}$ $\sim$ ${\rm (c)'}$ and
those matrices are transformed to each other by an orthogonal matrix.
\subsection{The new condition for the Fano operator}
Now we try to determine the Fano operator uniquely
by imposing
a similar condition to the condition Eq.(\ref{gnewconc}).

Instead of
the unitary operator $R_{\theta}$ in Eq.(\ref{gnewconc})
we use the unitary matrix $U$ which induces the transformation for 
matrices $P$ and $S$
\begin{eqnarray}
P & {\rightarrow} & UPU^{-1}= a_pS^{\mu}P^{\nu}, \label{newP} \\ 
S & {\rightarrow} & USU^{-1}= a_sS^{\kappa}P^{\lambda}, \label{newS}
\end{eqnarray}
where integers $\mu$, $\nu$, $\kappa$, $\lambda$ and phase factors $a_p$, $a_s$
are determined such that the matrices $UPU^{-1}$ and $USU^{-1}$ 
satisfy the same conditions as matrices $S$ and  $P$ do.
From Eq.(\ref{commSP}), for integers $\mu$,$\nu$,$\kappa$ and $\lambda$, we get 
\begin{equation}
\kappa \nu-\mu \lambda = 1, 
\label{conint}
\end{equation}
and from $S^N=1$ and $P^N=1$, we have
\begin{eqnarray}
a_p &=& \omega^{(N-1)\mu \nu /2}, \label{phs1} \\
a_s &=& \omega^{(N-1)\kappa \lambda /2}, \label{phs2} 
\end{eqnarray}

Comparing Eqs.(\ref{newP}) and (\ref{newS})
with Eqs.(\ref{gTraeQ}) and (\ref{gTraeP}), 
we find the correspondence for the parameters of both transformations
\begin{eqnarray*}
A \rightarrow  \nu ,&\;\;\;& B \rightarrow -\mu , \\
C \rightarrow -\lambda ,&\;\;\;& D \rightarrow  \kappa . 
\end{eqnarray*}
Thus, on the Fano operator on the lattice, we impose the new condition
\begin{equation}
U\hat{\Delta}(q,p)U^{-1}=\hat{\Delta}(\kappa q+\mu p,\nu p+\lambda q).
\label{newfrel}
\end{equation}
\subsection{the Fano operator under the new condition}
First, we find the condition for the coefficient $a(q,p;n,m)$
in Eq.(\ref{expdFano}) which is equivalent to Eq.(\ref{newfrel}).

From Eq.(\ref{expdFano}), the LHS of Eq.(\ref{newfrel}) becomes
\begin{eqnarray}
U\hat{\Delta}(q,p)U^{-1}&=&\sum_{n,m \in Z_N}a(q,p;n,m)(USU^{-1})^n(UPU^{-1})^m 
\nonumber \\
&=&\sum_{n,m \in Z_N}a(q,p;n,m)a_s^na_p^m(S^{\kappa }P^{\lambda })^n(S^{\mu }P^{\nu })^m 
\nonumber \\
&=&\sum_{n,m \in Z_N}a(q,p;n,m)\omega^{\phi(n,m)}S^{\kappa n+\mu m}P^{\lambda n+\nu m},
\label{RHSncon} 
\end{eqnarray}
where $\phi(n,m)$ is given by
\[
\phi(n,m)=\frac{1}{2}\{n(N-n)\kappa \lambda +m(N-m)\mu \nu \}-nm\mu \lambda, 
\]
and we can show that $\omega^{\phi(n,m)}$ is a periodic function
\[
\omega^{\phi(n+aN,m+bN)}=\omega^{\phi(n,m)}\;\; (a,b={\rm integer}).
\]
Instead of the summation with respect to $n$ and $m$, we make the summation 
with respect to $n'$ and $m'$ which are given by
\begin{equation}
\left\{
   \begin{array}{l}
     n'=\kappa n+\mu m \\
     m'=\lambda n+\nu m
\end{array}
\right..
\label{valtra}
\end{equation}
Owing to the condition Eq.(\ref{commSP}), we can solve Eq.(\ref{valtra}) for
$n$ and $m$,
\begin{equation}
\left\{
   \begin{array}{l}
     n=\nu n'-\mu m' \\
     m=-\lambda n'+\kappa m'
\end{array}
\right.. 
\label{invvltra}
\end{equation}
Substituting these equations into $\phi(n,m)$
we obtain 
\begin{equation}
\phi(\nu n'-\mu m',-\lambda n'+\kappa m')=\phi'(n',m')+\delta+N\times ({\rm integer}).
\end{equation}
where $\phi'(n',m')$ and $\delta$ are given by
\begin{eqnarray}
\phi'(n',m')&=&\frac{1}{2}\{\nu \lambda n'(N-n')+\mu \kappa m'(N-m')\}
+\mu \lambda n'm', \label{newph} \\
\delta&=&N\left\{\frac{\nu \lambda (\kappa -\mu -1)}{2}n'
+\frac{\kappa \mu (\nu -\lambda -1)}{2}m'\right\}.
\label{residual}
\end{eqnarray}
We can see that the values $\nu \lambda (\kappa -\mu -1)$ and $\kappa \mu (\nu -\lambda -1)$
are even under the condition Eq.(\ref{conint}) as follows.
We have two cases where $\nu \lambda (\kappa -\mu -1)$ is odd. The one is the case
where $\kappa $,$\lambda $, $\mu $ and $\nu $ are odd, the other is the case
where $\lambda $ and $\nu $ are odd and $\kappa $ and $\mu $ are even.
For both cases the value ${\kappa \nu-\mu \lambda}$ becomes even and
the condition Eq.(\ref{conint}) is not
satisfied. Similarly $\kappa \mu (\nu -\lambda -1)$
in the second term is even, so that we can replace the function
$\omega^{\phi(\nu n'-\mu m',-\lambda n'+\kappa m')}$ by $\omega^{\phi'(n',m')}$.
It can be easily checked that the function $\omega^{\phi'(n',m')}$ is a periodic function of  
period $N$,
\[
\omega^{\phi'(n'+aN,m'+bN)}=\omega^{\phi'(n',m')}\;\;(a,b={\rm integer}),
\]
and that this function is a function on $Z_N \times Z_N$.
As the coefficients of the linear transformation Eq.(\ref{valtra}) and the
inverse transformation Eq.(\ref{invvltra}) are integer,
the map defined by
Eq.(\ref{valtra}) has one-to-one correspondence on $Z_N \times Z_N$.
So we can make summation by $n'$ and $m'$ over
$Z_N \times Z_N$ instead of $n$ and $m$ over the same region:
\begin{equation}
U\hat{\Delta}(q,p)U^{-1}=
\sum_{n',m' \in Z_N}a(q,p;\nu n'-\mu m',-\lambda n'+\kappa m')
\omega^{\phi'(n',m')} S^{n'}P^{m'}
\label{RHSfin}
\end{equation}
Thus, the condition (\ref{newfrel}) in terms of $a(q,p;n,m)$ is written by
\begin{equation}
a(q,p;\nu n-\mu m,-\lambda n+\kappa m)
\omega^{\phi'(n,m)}=a(\kappa q+\mu p,\nu p+\lambda q;n,m).
\label{newfrel2}
\end{equation}
For the Fourier transformed coefficient $\tilde{a}(s,t;n,m)$,
this condition becomes
\begin{equation}
\tilde{a}(\nu s+\lambda t,\mu s+\kappa t;n,m)=
\omega^{\phi'(n,m)}\tilde{a}(s,t;\nu n-\mu m,-\lambda n+\kappa m).
\label{frel4fa}
\end{equation}
Replacing variables $s$ and $t$ by $\kappa s-\lambda t$ and $\nu t-\mu s$,
respectively, we get
\begin{equation}
\tilde{a}(s,t;n,m)
=\omega^{\phi'(n,m)}
\tilde{a}(\kappa s-\lambda t, \nu t-\mu s,t;\nu n-\mu m,-\lambda n+\kappa m).
\label{newfre3}
\end{equation}
As the coefficient $\tilde{a}(s,t;n,m)$ is a function on 
$Z_N \times Z_N \times Z_N \times Z_N$, if we can choose the integers
$\kappa$ and $\lambda$ such that 
\begin{equation}
\kappa s-\lambda t={\rm integer} \times N,
\label{con4kl}
\end{equation}
owing to condition ${\rm (a)'}$, we can obtain $\tilde{a}(s,t;n,m)$.
Thus, we can determine the coefficient $\tilde{a}(s,t;n,m)$ for any integers $s$ and $t$,
uniquely.

Now, we try to get an explicit form of $\tilde{a}(s,t;n,m)$.  We choose $\sigma$ and $\tau$ as
$\lambda$ and $\kappa$ respectively,
which are defined by,
\begin{eqnarray}
s &=& \xi\sigma+N\left[\frac{s}{N} \right]
(0 \leq \sigma \leq N-1), \label{vals} \\
t &=& \xi\tau+N\left[\frac{t}{N} \right]
(0 \leq \tau  \leq N-1), \label{valt}
\end{eqnarray}
where $\xi$ is the greatest common divisor of $\xi\sigma$ and
$\xi\tau$, namely, $\sigma$ and $\tau$ are relatively prime to each other.
If we choose $\sigma$
and $\tau$ as $\kappa$ and $\lambda$, respectively,
from number theory, it is clear that there exist the integers $\mu$ and 
$\nu$ satisfying the equation (\ref{conint}).
Then we obtain
\begin{eqnarray}
&& \tilde{a}(s,t;n,m)
=\omega^{\phi'(n,m)}
\tilde{a}(0, \xi;\nu n-\mu m,-\lambda n+\kappa m) \nonumber \\
&&\;\;\;\;\;=
\frac{1}{N^2}\omega^{\phi'(n,m)}
\delta^{(N)}_{s,m}\delta^{(N)}_{t,n} \nonumber \\
&&\;\;\;\;\;=
\frac{1}{N^2}\omega^{\phi'(t,s)}
\delta^{(N)}_{s,m}\delta^{(N)}_{t,n}.
\label{finfa1}
\end{eqnarray}
From Eq.(\ref{newph}), the $\phi'(s,t)$ with the choice of $\kappa$ and
$\lambda$ becomes
\[
{\phi'(s,t)}
=\frac{1}{2}\{N\xi \sigma \tau (\nu+\mu+1)-N\xi\sigma \tau -\xi^2 \sigma \tau \}.
\]
Using the same discussion as we did before Eq.(\ref{RHSfin}),
we can show that the first term in the above equation is
$N$ multiplied by an integer, and we have
\begin{equation}
\omega^{\phi'(s,t)}=\omega^{-\frac{1}{2}(N\xi\sigma \tau + \xi^2 \sigma \tau)}.
\end{equation}
We can replace $\xi$ by $\xi^2$ in the first term in the phase factor
$\omega^{\phi'(s,t)}$,
since $\omega^{N/2}=-1$ and $\omega^{N\xi/2}$ is equal to $\omega^{N\xi^2/2}$. 
We get
\begin{equation}
\tilde{a}(s,t;n,m)=\frac{1}{N^2}
\omega^{-(s-N\left[\frac{s}{N}\right])(t-N\left[\frac{t}{N}\right])(N+1)/2}
\delta^{(N)}_{s,m}\delta^{(N)}_{t,n}.
\label{finfra}
\end{equation}

On the process to get this solution, we imposed the condition (\ref{frel4fa})
for the 
integers $\kappa$, $\lambda$, $\mu$ and $\nu$ corresponding to 
the only one line which passes through the fixed point $(s,t)$. So it is not clear that 
this solution satisfies the condition (\ref{frel4fa}) for the four integers corresponding to
other lines. We have to
investigate whether the four conditions are satisfied by this solution or not.

When $N$ is odd, we can drop the term proportional to
$N$ in the phase factor as $N+1$ is even, and the solution becomes simple
\begin{equation}
\tilde{a}(s,t;n,m)=\frac{1}{N^2}
\omega^{-st(N+1)/2}
\delta^{(N)}_{s,m}\delta^{(N)}_{t,n},
\label{finfra2}
\end{equation}
which is equivalent to the solution given by Cohendet {\it et al.}\cite{cohendet} 
\begin{equation}
\tilde{a}(s,t;n,m)
   =\left\{\begin{array}{cc}
    \frac{1}{N^2} \omega^{-\frac{n    m}{2}}
                  \delta^{(N)}_{s,m}\delta^{(N)}_{t,n} & (n={\rm even}) \\
     \frac{1}{N^2} \omega^{-\frac{(n+N)m}{2}}
                  \delta^{(N)}_{s,m}\delta^{(N)}_{t,n}& (n={\rm odd })
                \end{array} \right..
\label{paperWa}
\end{equation}
It is proved that this solution satisfies conditions ${\rm (a)'}\sim{\rm (c)'}$
and we can easily check that the coefficient $\tilde{a}(s,t;n,m)$ satisfies the condition Eq.(\ref{frel4fa}). 

However, when $N$ is even, the solution Eq.(\ref{finfra}) does not satisfy
the condition ${\rm (b)'}$  because
\[
\left[\frac{-s}{N} \right] \neq -\left[\frac{s}{N} \right].
\]

Thus there dose not exist the Fano operator which satisfies original three conditions
${\rm (a)'}$ $\sim$ ${\rm (c)'}$ and the new condition Eq.(\ref{newfrel}). 
\section{summary and discussion}
In this note, in order to determine the Wigner function uniquely, we proposed a new condition.
For a continuous quantum system, the condition Eq.(\ref{gnewconc}) is the almost same
condition as Bertrand and Bertrand presented \cite{BertTomo}. 
However, this condition is simpler than their condition and can be easily extended to the system 
with a lattice phase space.

If the Fano operator satisfies the condition Eq.(\ref{gnewconc}), we can easily estimate the integration along a tilted line, say $Cq + Dp=p_0$, and
we have
\begin{eqnarray*}
&&\int^{\infty}_{-\infty}\hat{\Delta}(q=Dr-Bp_0,p=Ap_0-Cr)dr \\
&&\;\;\;\;\;\;=\int^{\infty}_{-\infty}
R_{\alpha \beta}\hat{\Delta}(p_0,r)R_{\alpha \beta}^{-1}dr \\
&&\;\;\;\;\;\;=R_{\alpha \beta}\left[
\int^{\infty}_{-\infty}\hat{\Delta}(p_0,r)dr\right]R_{\alpha \beta}^{-1} \\
&&\;\;\;\;\:=R_{\alpha \beta}\mid p_0 \rangle \langle p_0 \mid
R_{\alpha \beta}^{-1}.
\end{eqnarray*}
Clearly, the state $R_{\alpha \beta}\mid p \rangle$ is an eigenstate
of the Hermite operator $\hat{P}_G$,
\[
\hat{P}_G=R_{\alpha \beta}\hat{P}R_{\alpha \beta}^{-1}=C\hat{Q}+D\hat{P}.
\]
The integration of the Wigner function along this tilted line 
gives the distribution function over the eigenvalues of
the operator $C\hat{Q}+D\hat{P}$. Thus the Wigner function gives a
correct marginal distribution along a tilted line and 
this is important property  for quantum tomography.

On the analogy of a continuous quantum system, we presented the condition Eq.(\ref{newfrel})  
for the system whose phase space is a lattice with $N^2$ sites. Under this condition, for the case where 
$N$ is odd, we found out the same solution that Cohendet {\it et al.}
did\cite{cohendet}, and for the case where $N$ is even we showed that there does not exist the solution.

For the Fano operator on a lattice which satisfies the condition
Eq.(\ref{newfrel}), the summation 
of the Fano operator over the sites on the line $\kappa p-\lambda q=p_0$ becomes
\[
\sum_{r \in Z_N} \hat{\Delta}(q=\kappa r+ \mu p_0, p=\nu p_0 +\lambda r)
=U \mid p_0 \rangle \langle p_0 \mid U^{-1},
\]
which is the projection operator to an eigenvector of $USU^{-1}$ as is seen from Eqs.(\ref{eigens})
and (\ref{newS}),
so that we can get the correct marginal distribution on a tilted line for this system. 
 
Finally, we point out that the Fano operator is determined uniquely from
only two conditions, namely, condition (a) and Eq.(\ref{gnewconc}) for 
a continuous quantum system and  condition ${\rm (a)'}$ and Eq.(\ref{newfrel})
for a lattice.


\end{document}